\documentclass[journal]{IEEEtran}

\usepackage{cite}

\ifCLASSINFOpdf
  \usepackage[pdftex]{graphicx}
\else
\fi

\usepackage{amssymb}
\usepackage{xcolor}
\usepackage{amsmath}
\usepackage{epsfig}
\usepackage{comment}
\usepackage{subfigure}
\usepackage{subcaption}
\usepackage{multirow}
\usepackage{makecell}
\usepackage{array}
\usepackage[lofdepth,lotdepth]{subfig}
\begin{document}

\title{Experimental Performances of mmWave RIS-assisted 5G-Advanced Wireless Deployments in Urban Environments}

\IEEEpubid{\makebox[\columnwidth]{{XXX-Y-ZZZZ-WWWW-T/UU/\$31.00 ©2024 IEEE}\hfill}
\hspace{\columnsep}\makebox[\columnwidth]{}}

\author{\IEEEauthorblockN{Ahmet Faruk COŞKUN\IEEEauthorrefmark{1}, Alper Tolga KOCAOĞLU\IEEEauthorrefmark{2}, Emre ARSLAN\IEEEauthorrefmark{1}, Zehra YİĞİT\IEEEauthorrefmark{1}, Samed KEŞİR\IEEEauthorrefmark{1}, \\Batuhan KAPLAN\IEEEauthorrefmark{1}, Jianwu DOU\IEEEauthorrefmark{3}, Yijun CUI\IEEEauthorrefmark{3}} \\
\IEEEauthorblockA{\IEEEauthorrefmark{1}6GEN Laboratory, Next-Generation R\&D, Network Technologies, Turkcell, {\.{I}}stanbul, Turkiye}

\IEEEauthorblockA{\IEEEauthorrefmark{2}Access Network Architecture, Network Technologies, Turkcell, {\.{I}}stanbul, Turkiye\\ \IEEEauthorrefmark{3}ZTE Corporation, Shenzhen, 518055 China,\\ Emails: {\{coskun.ahmet, alper.kocaoglu, emre.arslan, zehra.yigit, samed.kesir, batuhan.kaplan\}@turkcell.com.tr,}\\{\{dou.jianwu, cui.yijun\}@zte.com.cn}}}

\maketitle

\begin{abstract}
Reconfigurable intelligent surface (RIS) has emerged as a groundbreaking technology for 6G wireless communication networks, enabling cost-effective control over wireless propagation environment. By dynamically manipulating its codebook so as to deflect the direction of the reflected electromagnetic wave, RIS can achieve enhanced signal quality, extended coverage, and interference mitigation. This study presents experimental performance of ZTE Dynamic 2.0 RIS products through a series of real-world tests conducted on Turkcell's millimeter-wave (mmWave) testbed. The evaluation involves network coverage extension in urban areas, multi-user efficiency, and the integration of virtual reality technology to support immersive applications in next-generation 6G networks. Through a comprehensive measurement-based analysis, the performance of the RIS product is demonstrated, highlighting its potential to address critical challenges in mmWave communications and to enable advanced 6G use cases.
\end{abstract}

\begin{IEEEkeywords}
5G-Advanced, 6G, reconfigurable intelligent surfaces (RISs), millimeter wave (mmWave) communications
\end{IEEEkeywords}

\section{Introduction}
Reconfigurable Intelligent Surface (RIS) technology has emerged as a pivotal enabler for future 6G communication networks due to its unique capability to flexibly manipulate the properties of reflected signals. This functionality promises significant advancements in signal quality, coverage extension, and interference management, making RIS a cornerstone in the evolution of next-generation wireless communication. While theoretical studies have demonstrated the potential of RIS in various scenarios, the transition from theoretical frameworks to practical deployment remains underexplored. 
Experimental studies are indispensable as they provide crucial insights into RIS’s operational characteristics under realistic conditions. Currently, the literature lacks extensive experimental studies on RIS performance in realistic environments. The few existing investigations often rely on simplified prototypes or constrained testing conditions, which fail to capture the complexities of real-world deployments. For RIS to be effectively integrated into commercial networks, a comprehensive understanding of its practical limitations, deployment challenges, and performance metrics is imperative. Practical analyses, incorporating realistic configurations and advanced prototypes/commercial test devices, are necessary to validate RIS’s utility and scalability in live networks.

Globally, applied research on RIS technology is predominantly conducted using testbeds and development boards \cite{trichopoulos2022design, li2022path, mazloum2023impact}. Several studies have developed accurate path loss models for RIS, accounting for unit cell directivity characteristics \cite{tang2022path}, \cite{Miao2025Modeling}. Another study addressed RIS-assisted communication, considering RIS size and near-field/far-field effects, with experimental validation in a microwave anechoic chamber \cite{tang2020wireless}. Besides, research on channel characterization of RIS-assisted wireless links has addressed various aspects of RIS-based channels, including reflection, transmission, and multipath fading mitigation \cite{ huang2022reconfigurable}. These studies have classified RIS-based channel measurements by frequency bands, scenarios, and system configurations, providing a comprehensive overview of channel characteristics and modeling efforts \cite{tang2021channel, mudonhi2021ris, xilong2021trials, sang2024measurement}.

In addition, RIS tests conducted with product-level infrastructures are rarely encountered \cite{karakucs2024indoor}. Experimental studies conducted in different indoor/outdoor environments and across several frequency bands which are $2.6$ GHz, $4.9$ GHz, and $26$ GHz confirm the effectiveness of RISs in enhancing signal strength and data rates \cite{ZTEFieldTrial}. Also, in \cite{sang2022coverage}, RIS deployments in current 5G networks were performed to analyze coverage enhancement for various shadow zone scenarios at $2.6$ GHz. These experimental efforts highlight the practical benefits and challenges of deploying RIS in wireless networks.

Despite providing solid test trials, \cite{ ZTEFieldTrial} and \cite{sang2022coverage} lack to put analyze a commercial-like RIS product 
which should be integrated to the cellular network in terms of a control interface. In this paper, the seamless integration of the RIS into the cellular network is accomplished to serve moving user continuously, where the best beam either from base station to user or RIS to user is selected in the network. Hence, the future direction issues 
mentioned in \cite{sang2022coverage} are addressed by conducting a detailed measurement campaign to analyze the integration of mmWave RIS products into a cellular network. Also, in this paper, the  mmwave RIS providing required low latency and high throughput for virtual reality (VR) application is examined. The remainder of this manuscript is structured as follows: Section II and Section III describes the experimental setup and presents the test results and analysis, respectively. Furthermore, Section IV includes critical discussions and future directions. Finally, Section V concludes the paper.

\section{Experimental Setup: RIS-Assisted 5G-A Wireless Networks}
This section introduces the network architecture, the radio access network (RAN) components, the user equipment, and the methodology of the measurement campaign

\subsection{Network Architecture and Equipment Specifications}
For the deployments planned at two different Turkcell sites to evaluate the coverage and signal quality enhancements of 5G mmWave RIS-assisted wireless networks, RAN side utilizes an n258-band mmWave active antenna unit (AAU) connected to a baseband unit (BBU), and \emph{NodeEngine} solution \cite{ZTENodeEngine} which is dedicated to build a simple but intelligent 5G private network. For the signal transmission period, a total of $400$ MHz cell bandwidth is configured with $4$T$4$R multiple-input multiple-output support in the Standalone mode. 

%%%%%%%%%%%%%%%%%%%%%%%%%%%%%%%
%%% Kartal Scenario Footage  %%
%%%%%%%%%%%%%%%%%%%%%%%%%%%%%%%
\begin{figure}[h]
    \centering
    {\includegraphics[width=0.99\columnwidth]{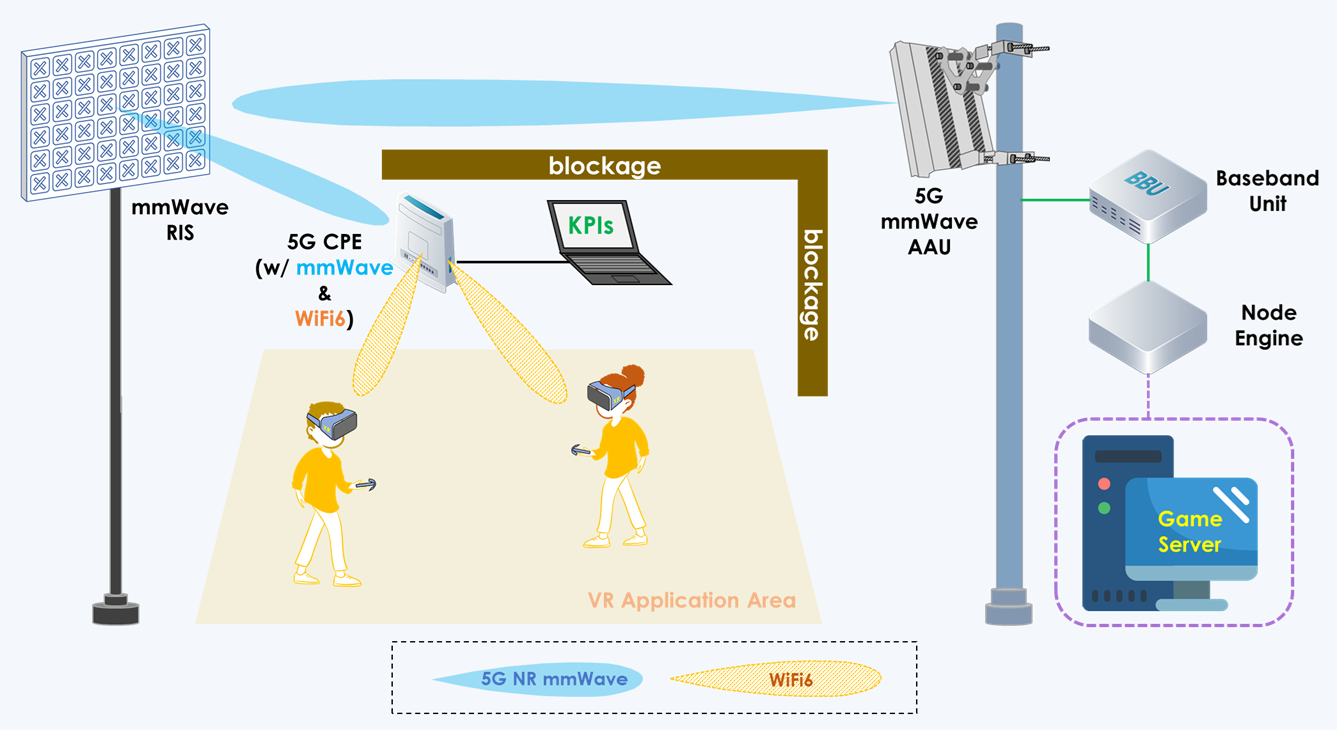}}
    \caption{Architectural diagram of RIS-assisted mmWave test setup}
    \label{fig:MainSystemArchitecture}
\end{figure}
%%%%%%%%%%%%%%%%%%%%%%%%%

As the new network component, the RIS Dynamic 2.0 YN6815 product of ZTE Corporation (hereinafter referred to as ZTE) has been utilized. This product contains two-state phase shifter elements arranged in a $132\times 132$ square array (a total of $17424$ elements). It is capable of forming predefined codebook-based beams to serve for a single user equipment.

During the tests conducted at Küçükyalı Plaza (i.e., introduced in Section III-B), a server computer functioning as a VR-based game server—responsible for processing, interpreting, and transferring data generated by VR headsets and handheld terminals—has been connected to the BBU and integrated into the 5G mmWave wireless network. This setup has enabled the mmWave transmission directed to the customer premises equipment (CPE) device via switching among the RIS codebooks. Then, the mmWave communications interface has been converted into a WiFi6 signal through the CPE’s mmWave/WiFi6 built-in modem capabilities, thereby ensuring seamless end-to-end communication for the VR devices.

\subsection{Methodology}
During the field tests, the RIS has been positioned within the direct line of sight (LoS) of the 5G AAU, acting as an anchor for the communication service provided by the AAU. Additionally, the mmWave CPE device, designated as the user terminal, has been moved along a trajectory where the majority of waypoints remain within the LoS of the RIS. Furthermore, as part of the pre-test network planning phase, Synchronization Signal Block (SSB) beams coordinated by the AAU via the configuration-loading codebooks of the RIS have been optimized to serve the CPE device on its trajectory. By leveraging RIS-generated SSBs with beamwidths adjustable in the $2^\circ$–$6^\circ$ range, the optimal SSBs for the CPE device have been identified, ensuring maximum communication quality.

The independently collected KPI records for both RIS-off and RIS-on cases along the specified trajectory have been used to match and compare RIS-free and RIS-assisted performance for each CPE waypoint. Although the recorded data include various parameters related to configuration, signal strength, and data rate, all time-stamped with coordinated universal time (UTC) labels, for ease of analysis, only the reference signal received power (RSRP), signal-to-noise plus interference ratio (SINR), uplink and downlink (UL/DL) throughput KPIs have been visualized in the subsequent sections of this study.

\section{Outdoor Test Scenarios and Results}
The extensive examination is split into two experimentation phases, first of which is planned as a validation and iterative trial to exhibit the achievements of RIS products better, whereas the second phase is more dedicated to the utilization of the RIS-assisted mmWave wireless network to serve for a latency-critical VR application. The described wireless network deployments have been set in two different premises of Turkcell (i.e., Kartal and Küçükyalı, respectively).

\subsection{Kartal Plaza Tests: Exhibition of Main RIS Achievements}
\begin{figure}[h]
    \centering
    {\includegraphics[width=0.99\columnwidth]{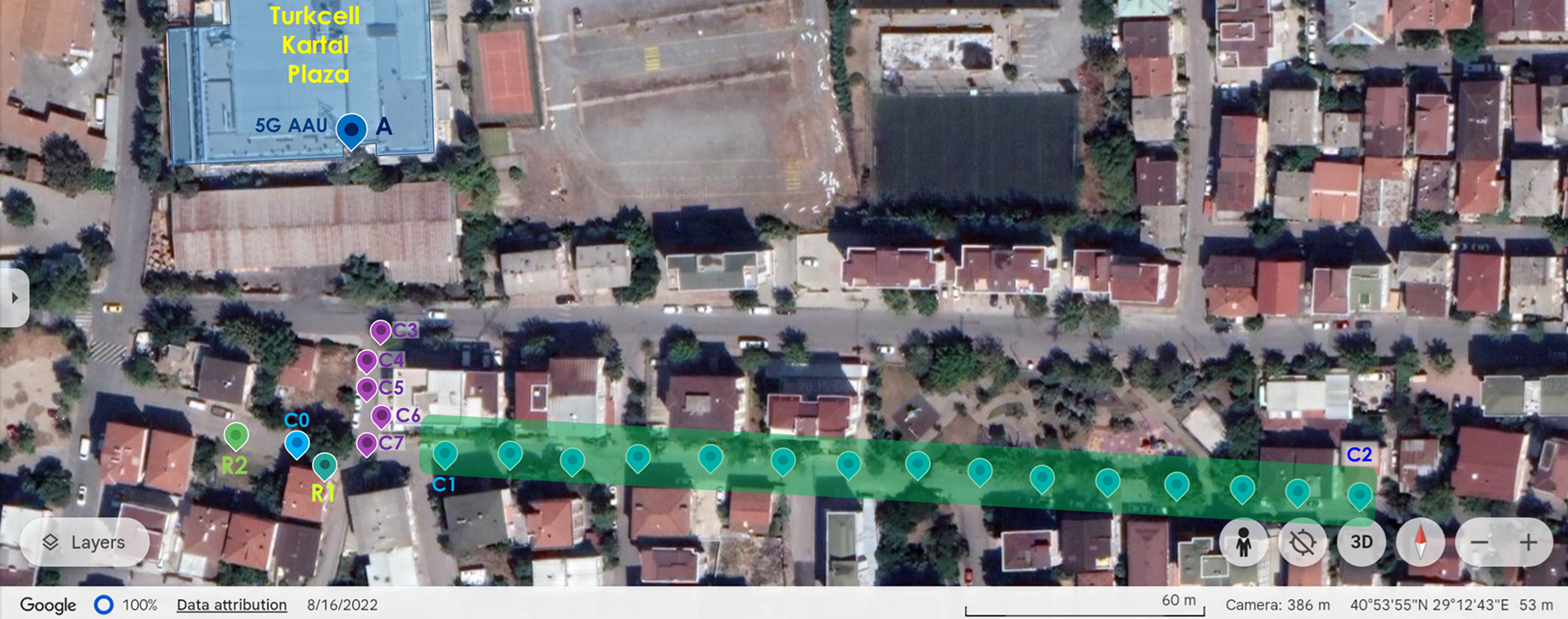}}
    \caption{Google Earth view of Kartal scenario}
    \label{fig:KartalPlazaScenarioA1}
\end{figure}

The RIS-assisted 5G-A test scenarios have been deployed in the vicinity of Kartal Plaza where the surrounding environment consists of urban residential houses and some industrial campuses. The satellite image of the demonstration scenario given in Fig. \ref{fig:KartalPlazaScenarioA1} includes the locations of the mmWave 5G-A AAU, RIS product, and the CPE waypoints. In Fig. \ref{fig:KartalPlazaScenarioA1}, the 5G-A AAU located at the rooftop of Kartal Plaza has been labeled as point \emph{A} with a dark blue marker. Two different RIS locations labeled as \emph{R1} and \emph{R2} shown with two different green color tones were considered during the field trials. Here, the light blue and purple points represent the CPE waypoints for two different drive tests and the fixed-point two-user test.

%%%%%%%%%%%%%%%%%%%%%%%%%%%%%%%
%%% Kartal Scenario Footage  %%
%%%%%%%%%%%%%%%%%%%%%%%%%%%%%%%
%%%%%%%%%%%%%%%%%%%%%%%%%

To comprehensively analyze and showcase the performance of mmWave RIS products, they have been installed at two different locations within the test area, each offering distinct dominance capabilities over the target street. At the first location (i.e., \emph{R1} in Fig. \ref{fig:KartalPlazaScenarioA1}), the RIS product has been placed on the balcony of a house to provide a clear LoS specifically to the entrance areas of \emph{Mavisu Street}. This setup represents a scenario where the signal strength and data transmission rate gains achieved by the RIS-based mmWave link are expected to reach their maximum potential. With the second RIS location (i.e., \emph{R2} in Fig. \ref{fig:KartalPlazaScenarioA1}), it is expected to cover a larger portion of potential CPE locations along \emph{Mavisu Street} more effectively. In this scenario, although a shadowing effect is expected due to the tree-lined area at the entrance of \emph{Mavisu Street} from the RIS perspective, it is expected to achieve maximum values in terms of the RIS-CPE range (maximum range scenario).

In a third scenario set up in the Kartal Plaza area, a two-user test setup was established for CPE locations marked in purple in Fig. \ref{fig:KartalPlazaScenarioA1}, along with another fixed CPE point (two-user scenario). Since the mmWave RIS Dynamic 2.0 product serves only a single user, in this scenario, one of the CPE devices (i.e., \emph{C0}) received service directly from the AAU, while the other was served via the RIS. The geographical information of the network equipment used in Kartal Plaza tests is given in Table \ref{table1:GeoInfo}. As seen, in addition to the coordinates of network nodes, the table includes $d_{RIS}$ which denotes the distance values of the AAU and CPE locations with respect to the utilized RIS location (i.e., \emph{R1} or \emph{R2}). The test results for these three scenarios are provided in the respective subsections.

%%%%%%%%%%%%%%%%%%%%%%%%%
\begin{table}[h]
  \centering
  \caption{\textsc{Geographical Information}}
  \label{table1:GeoInfo}
  \begin{tabular}{c c c c c}
    \hline
    {\bf{Equip.}} & {\bf{Latitude [$^\circ$]}} & {\bf{Longitude [$^\circ$]}} & {\bf{Altitude [m]}} & {\bf{$d_{RIS}$ [m]}}\\
    \hline
    \hline
    {\bf{AAU}} & & & & \\\hline\hline   
    {\emph{A}} & $40.899139$ & $29.210756$ & $88.7$ & $80.58$/$80.66$ \\\hline

    {\bf{RIS}} &  &  & &  \\\hline\hline
    {\emph{R1}} & $40.898419$ & $29.210645$ & $67.4$ & $-$ \\\hline
    {\emph{R2}} & $40.898478$ & $29.210360$ & $64.1$ & $-$ \\\hline

    {\bf{CPE}} &  &  & &  \\\hline\hline
    {\emph{C0}} & $40.898448$ & $29.210578$ & $57.8$ & $18.62$ \\\hline
    {\emph{C1}} & $40.898429$ & $29.210889$ & $59.2$ & $20.53$ \\\hline
    {\emph{C2}} & $40.898294$ & $29.213673$ & $58.8$ & $254.80$\\\hline
    {\emph{C3}} & $40.898691$ & $29.210755$ & $57.7$ & $40.77$\\\hline
    {\emph{C4}} & $40.898615$ & $29.210751$ & $57.7$ & $36.21$\\\hline
    {\emph{C5}} & $40.898531$ & $29.210741$ & $57.6$ & $32.55$\\\hline
    {\emph{C6}} & $40.898483$ & $29.210791$ & $57.6$ & $36.22$\\\hline
    {\emph{C7}} & $40.898421$ & $29.210730$ & $57.6$ & $31.73$\\\hline
  \end{tabular}
\end{table}

\subsubsection{Scenario of maximum RIS gains}
\begin{figure}
    \centering
    \subfigure[RSRP]
    {\includegraphics[width=0.98\columnwidth]{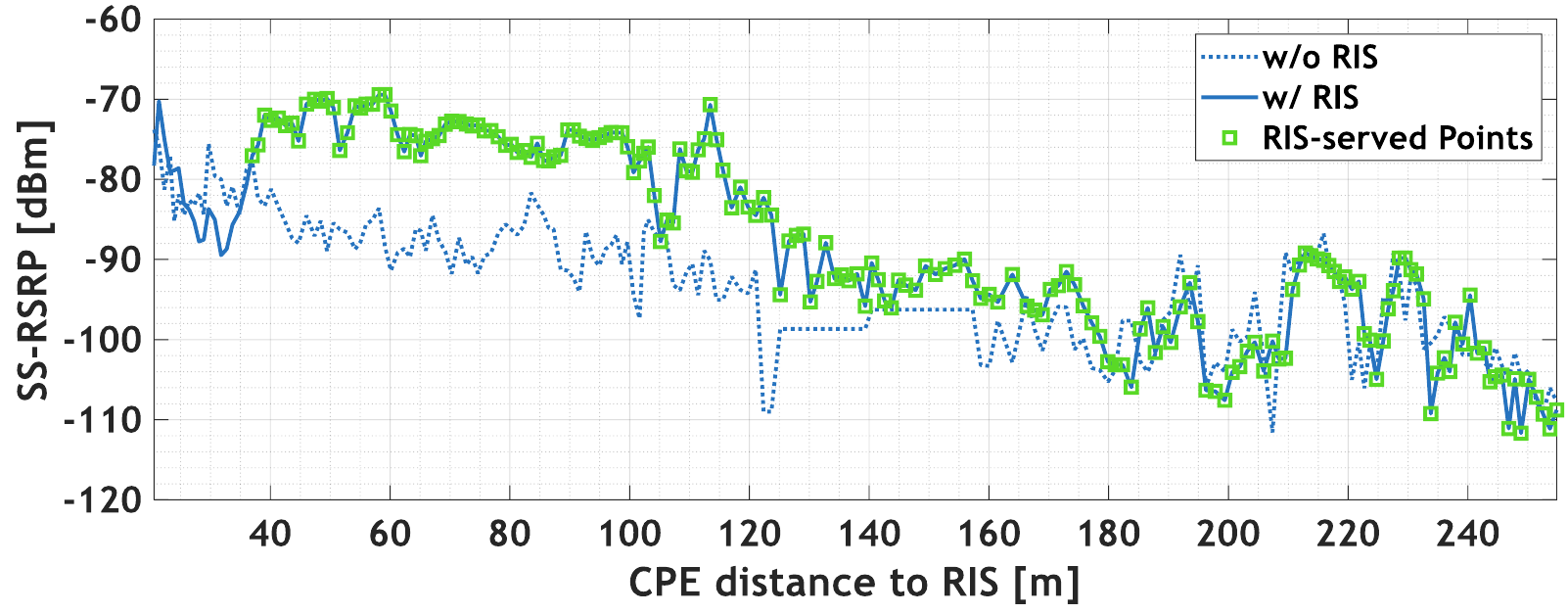}}
    \subfigure[SINR]
    {\includegraphics[width=0.98\columnwidth]{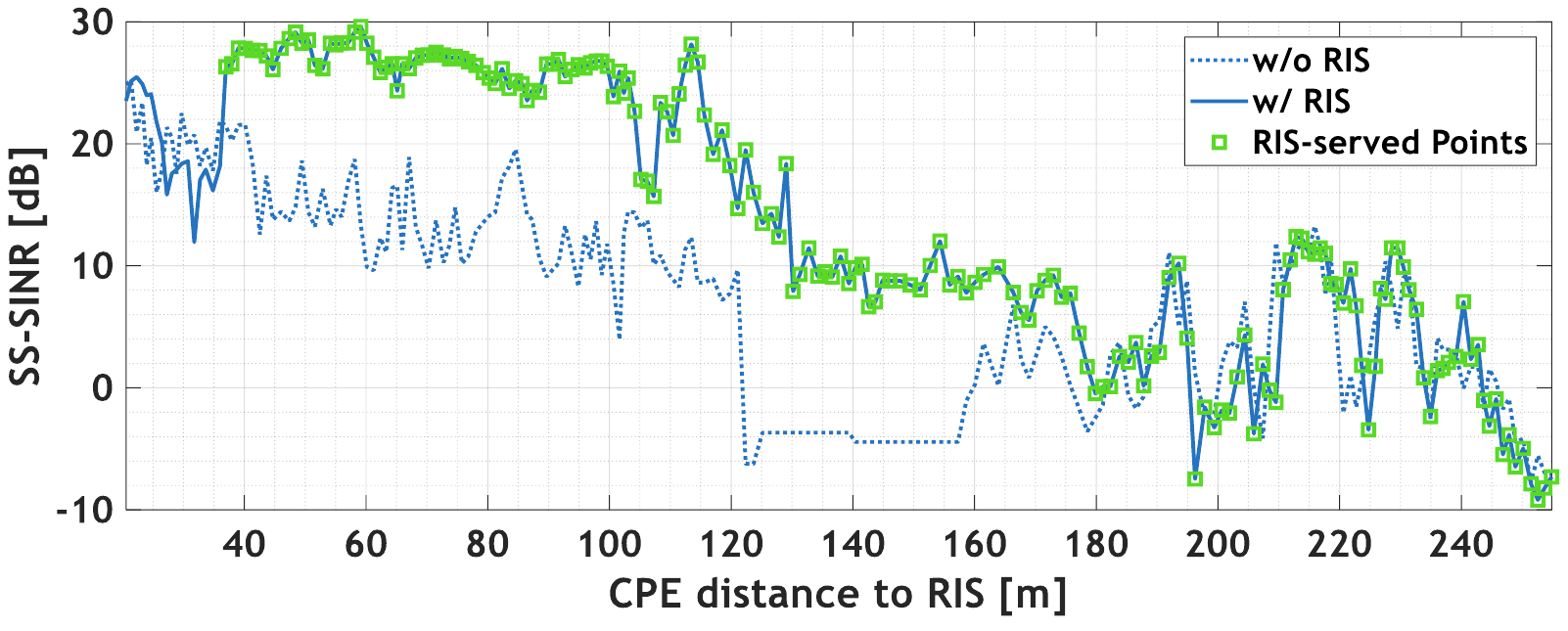}}
    \subfigure[UL Throughput]
    {\includegraphics[width=0.98\columnwidth]{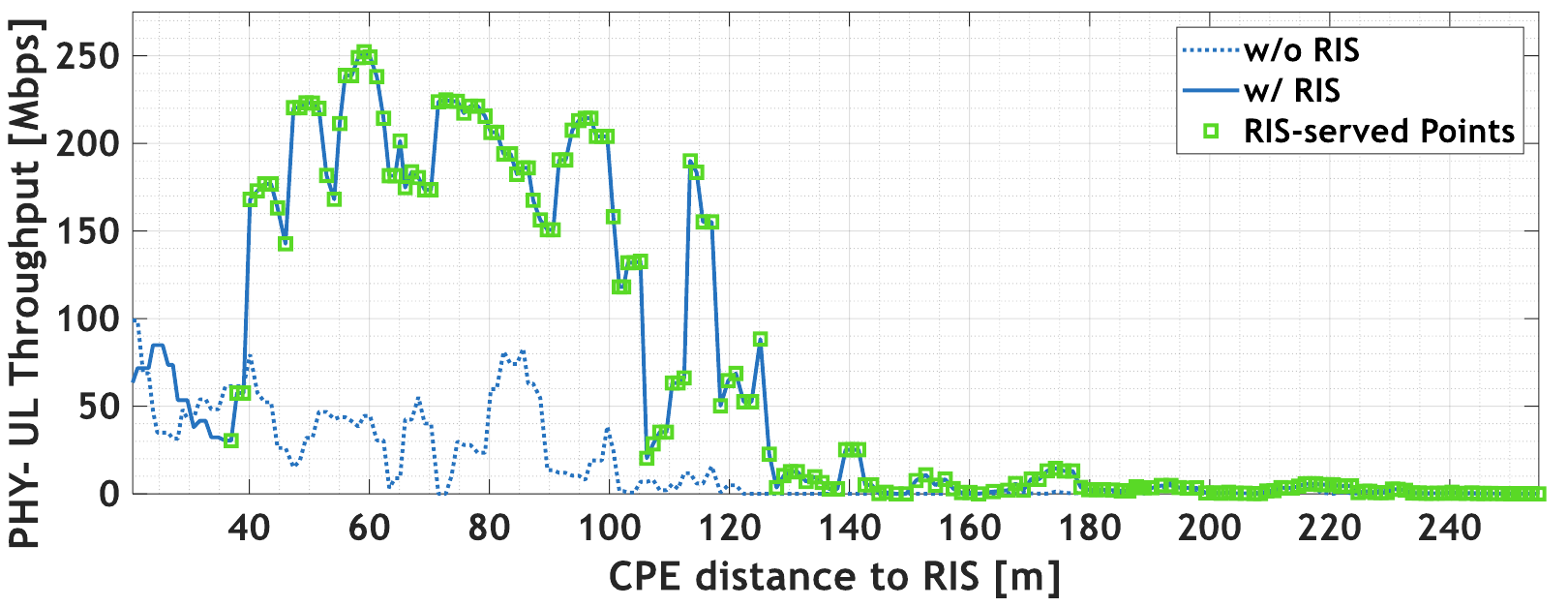}}
    \subfigure[DL Throughput]
    {\includegraphics[width=0.98\columnwidth]{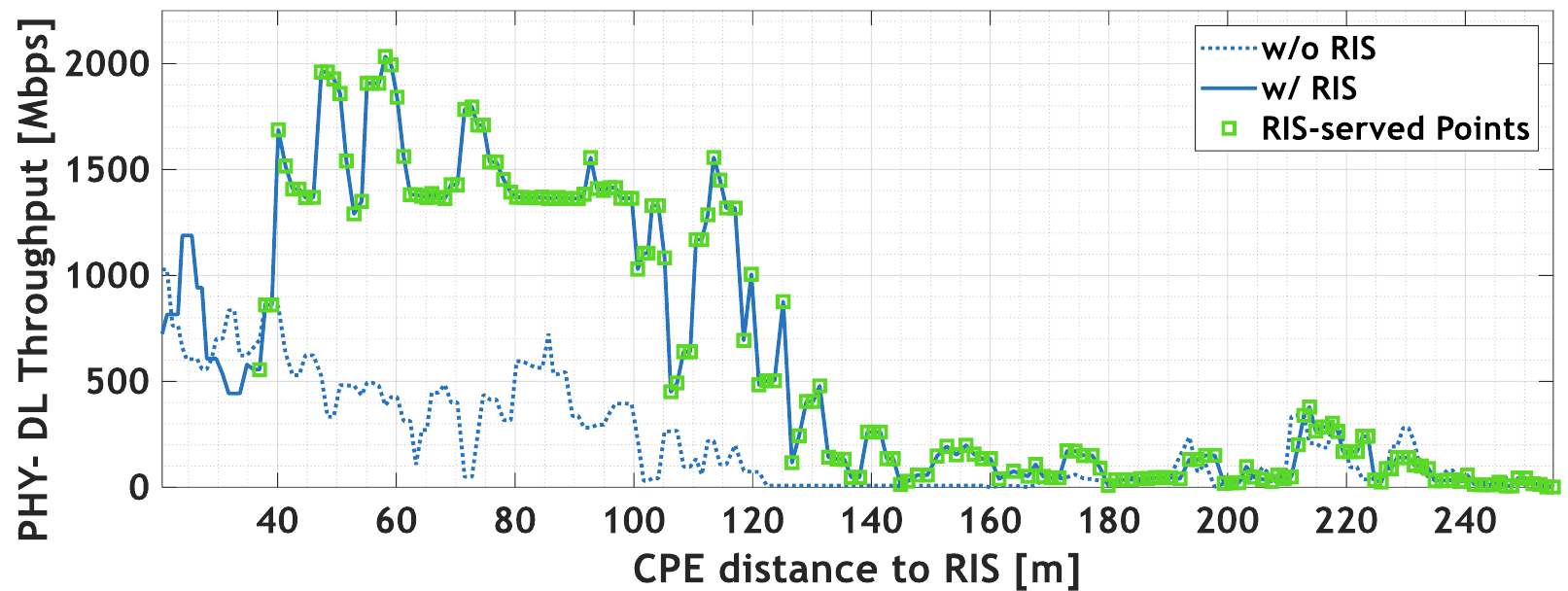}}    
    \caption{KPI records of the scenario: Maximum RIS-based gains}
    \label{fig:RISGainsForScenarioA}
\end{figure}

In this scenario, the communications quality of service (QoS) experienced by a mmWave CPE device mounted on a vehicle moving within the street is monitored, thanks to the RIS product placed on a building's second floor balcony (i.e., $7$ meters from ground) located at coordinate \emph{R1}, overlooking the western entrance of Mavisu Street. Here, KPIs representing the communications QoS are systematically recorded in the region between two positions indicated as \emph{C1} and \emph{C2} in Fig. \ref{fig:KartalPlazaScenarioA1}, then the RSRP, SINR, UL and DL throughput values have been depicted in Fig. \ref{fig:RISGainsForScenarioA}. By examining the KPI variations it can be clearly seen that the enhancements provided by the mmWave RIS product are solid for the CPE locations up to $110$ meters from RIS, and still exist but tend to decrease between this distance up to $180$ meters due to the horizontal curvature of the street. Within this $180$-meter serving corridor, the RIS-based advantage in the RF power-related metrics RSRP and SINR are typically about $20$ dB. On the other hand, the UL and DL data rates have been shown to be enhanced by $4$-folds typically with the activation of mmWave RIS.

%%%%%%%%%%%%%%%%%%%%%%%%%%%%%%%
%%% KPI Plots for max gains %%
%%%%%%%%%%%%%%%%%%%%%%%%%%%%%%%

\subsubsection{Scenario of maximum RIS assistance range}
\begin{figure}
    \centering
    \subfigure[RSRP]
    {\includegraphics[width=0.98\columnwidth]{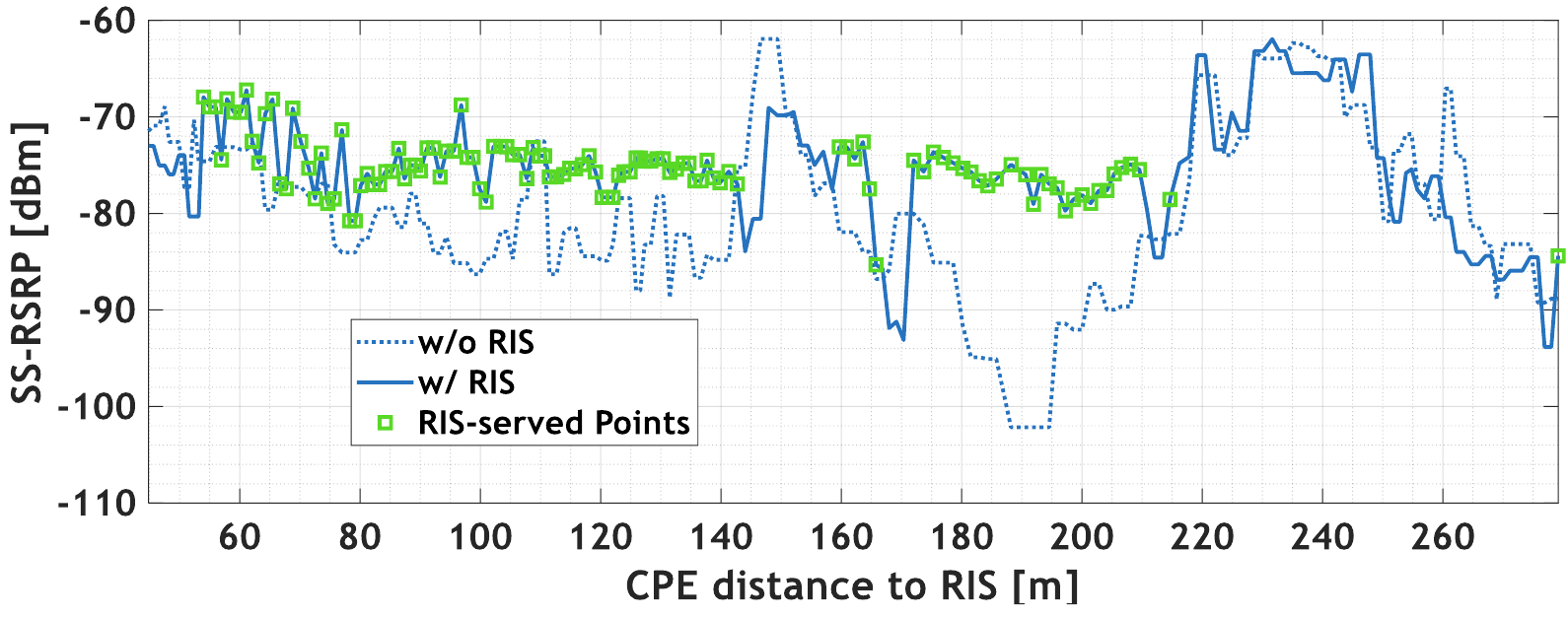}}
    \subfigure[SINR]
    {\includegraphics[width=0.98\columnwidth]{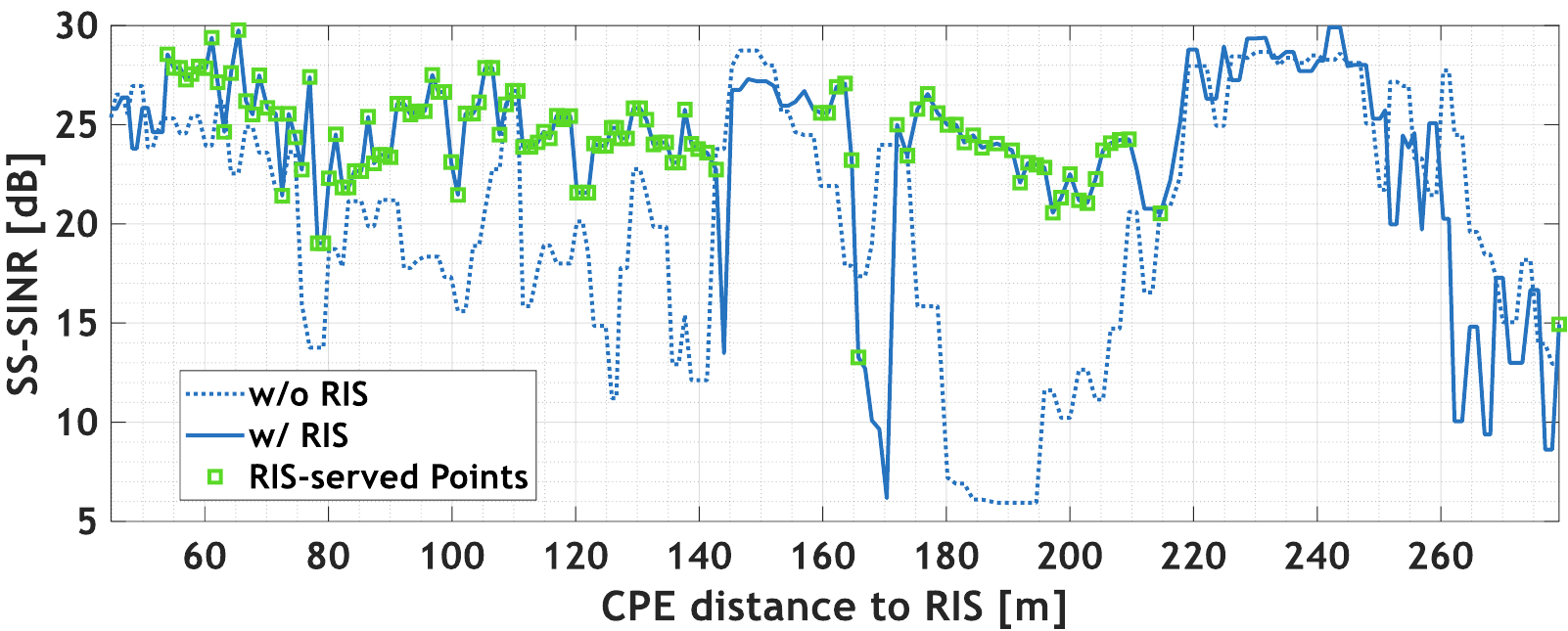}}
    \subfigure[UL Throughput]
    {\includegraphics[width=0.98\columnwidth]{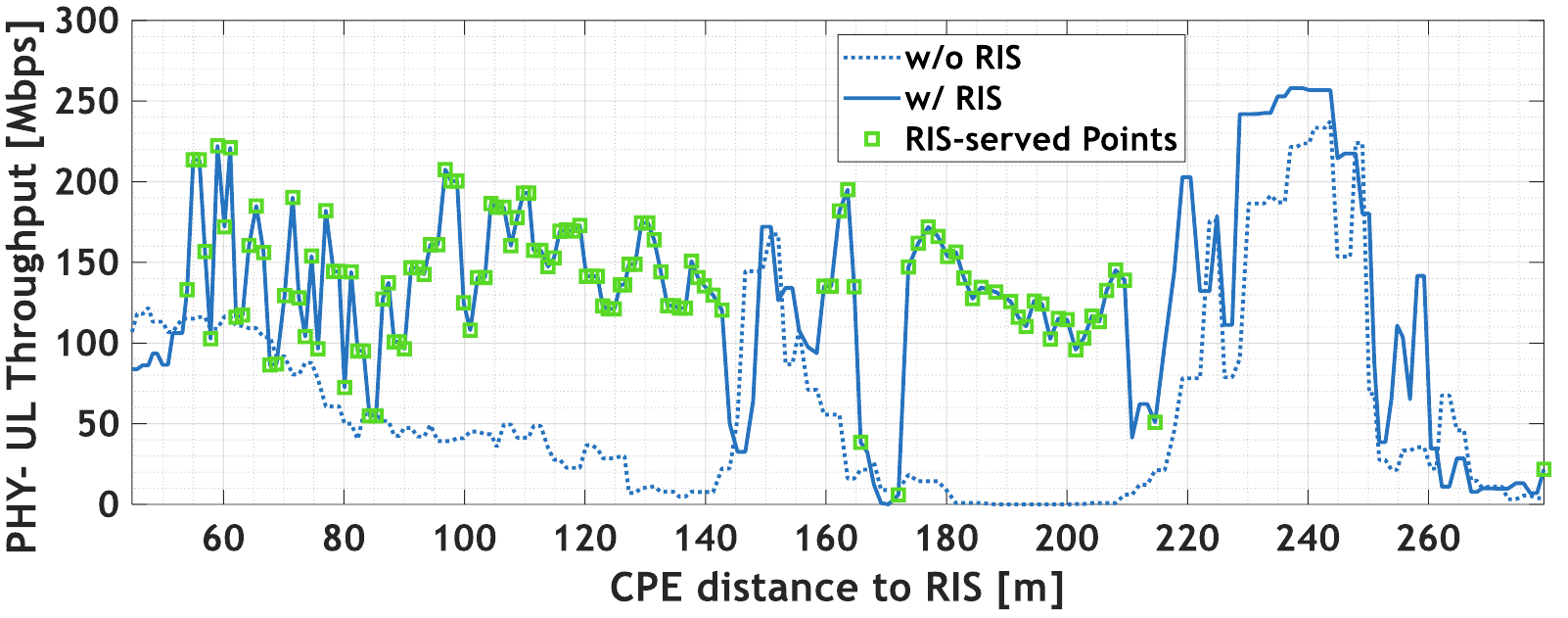}}
    \subfigure[DL Throughput]
    {\includegraphics[width=0.98\columnwidth]{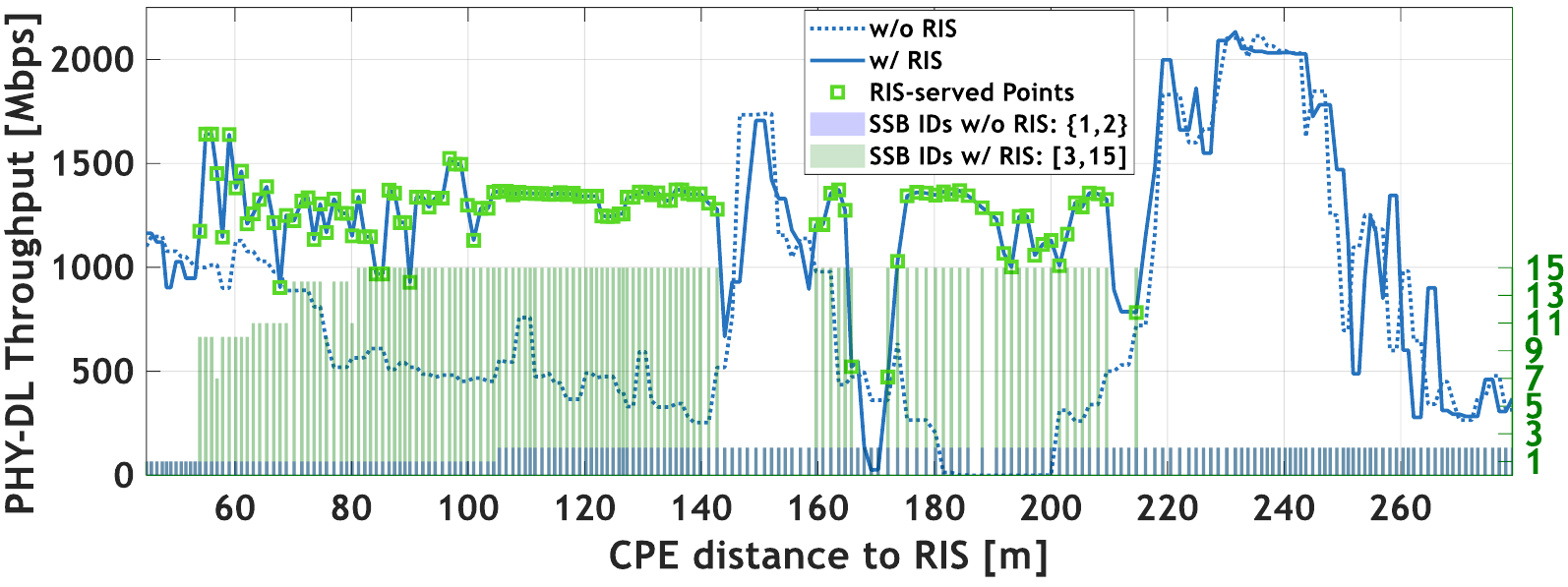}}    
    \caption{KPI records of the scenario: Maximum RIS-served range}
    \label{fig:RISGainsForScenarioB}
\end{figure}

Although RIS-induced KPI improvements reach almost peak levels, they tend to degrade and then vanish beyond the distance of approximately $110$ meters from the RIS due to the inability of the RIS at position \emph{R1} to address the southern curvature of the street. To extend this distance and to enhance the RIS-assisted mmWave RF signal penetration through the street, a more dominant RIS location overlooking the entire street (i.e., denoted as \emph{R2} in Fig. \ref{fig:KartalPlazaScenarioA1}) has been selected. For the new RIS location, the resulting KPI records have been processed and plotted in Fig. \ref{fig:RISGainsForScenarioB}. Here, it is observed that the maximum range, where the intense impact of RIS-induced gains on RSRP, SINR, and UL/DL throughput is evident, reaches approximately $210$ meters from the RIS. The improvements in communication QoS, which are clearly visible in Fig. \ref{fig:RISGainsForScenarioB}, are driven by the 5G AAU's ability to leverage the RIS as an anchor. This allows it to utilize the RIS SSBs for initial access, as well as the bi-statically provided data traffic beams via the RIS, and to perform beam tracking to adapt to changes in the CPE location. In order to showcase this tracking capability, the variation in the 'Serving SSB Beam Index' has been included as bar plots in Fig. \ref{fig:RISGainsForScenarioB}-(b).

%%%%%%%%%%%%%%%%%%%%%%%%%%%%%%%
%%% KPI Plots for max range %%
%%%%%%%%%%%%%%%%%%%%%%%%%%%%%%%
%%%%%%%%%%%%%%%%%%%%%%%%%
In the absence of RIS, the 5G AAU provides very limited mmWave communications quality as it registers the CPE over the SSB beams $1$ and $2$. However, with the activation of the RIS in the mmWave network, initial network access can be achieved through SSB beam indices within the interval $[3, 15]$, varying with respect to changes in the CPE location.

\subsubsection{Two-user scenario}
\begin{figure}[h]
    \centering
    {\includegraphics[width=0.99\columnwidth]{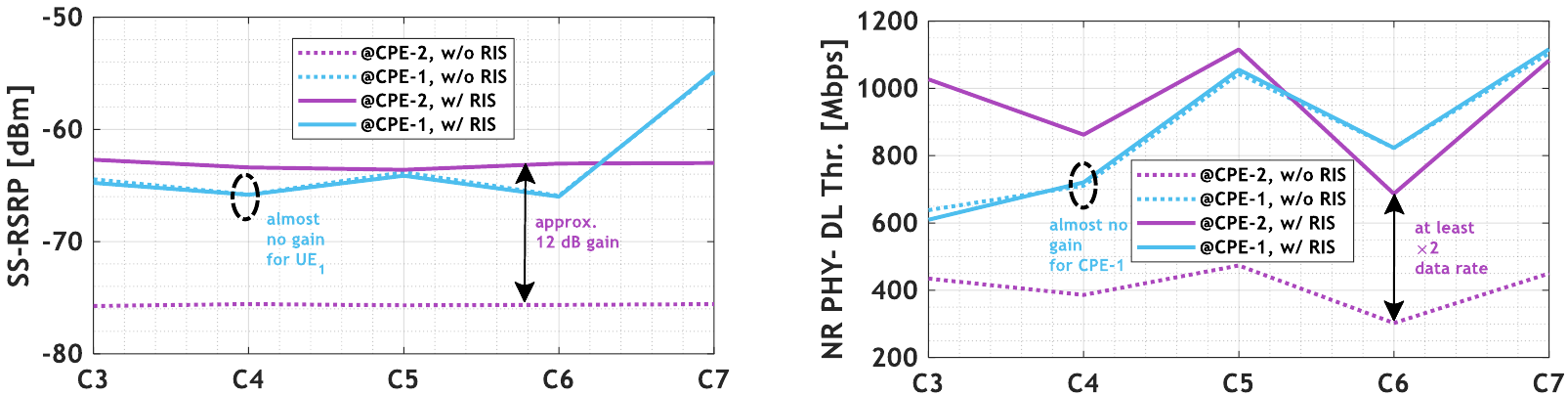}}
    \caption{Variation of RSRP and DL throughput for both CPEs}
    \label{fig:TwoUserKPIs}
\end{figure}

As stated in Section III-A, the Dynamic RIS 2.0 product can facilitate the registration of a single CPE and subsequently enables DL communications using 5G beamforming. For another CPE device located within the coverage area of the mmWave AAU, the registration and mmWave communication service will be conducted directly using the SSB and traffic beams generated by the AAU. In this scenario, as shown in Fig. \ref{fig:KartalPlazaScenarioA1}, the mmWave experiences of two CPE devices placed in close proximity were simultaneously analyzed. Here, the CPE device that directly receives the communication service from the AAU is referred to as CPE-1 (denoted as \emph{C0} in Fig. \ref{fig:KartalPlazaScenarioA1}), while the other (RIS-assisted) device is designated as CPE-2 (denoted as \emph{C3}-\emph{C7} in Fig. \ref{fig:KartalPlazaScenarioA1}). During the tests, when the RIS was inactive, both devices received broadband connectivity conventionally via the 5G AAU. However, when the RIS became active, the data flow between CPE-2 and the AAU occurred via the RIS, as it provided a more favorable data traffic pathway. While CPE-1 was kept stationary, KPI measurements were taken for five different positions of CPE-2. The recorded RSRP and DL throughput values are depicted in Fig. \ref{fig:TwoUserKPIs} with respect to CPE-2 locations. 
%%%%%%%%%%%%%%%%%%%%%%%%%%%%%%%
%% KPIs for Two-User Scenario %
%%%%%%%%%%%%%%%%%%%%%%%%%%%%%%%
%%%%%%%%%%%%%%%%%%%%%%%%%
When examining the power- and data rate-based KPIs observed on both devices in Fig. \ref{fig:TwoUserKPIs}, it is evident that the presence of the RIS has no effect on CPE-1, whereas the experience of CPE-2 has been enhanced as the RIS starts to assist the network. The depicted KPIs clearly validate the considerable performance boost of RIS assistance and also exhibit their offloading capabilities in mmWave networks. 

\subsection{Küçükyalı Plaza Tests: Integration of VR-based end user application}
\begin{figure}[h]
    \centering
    {\includegraphics[width=0.99\columnwidth]{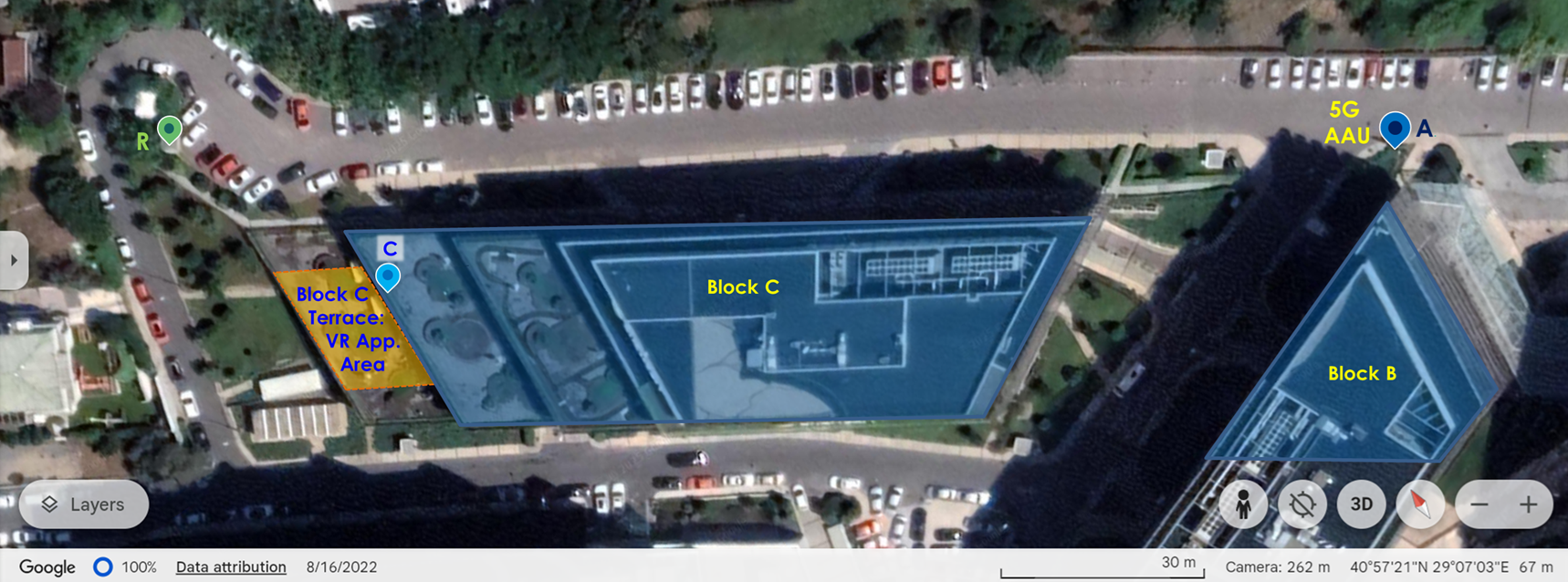}}
    \caption{Google Earth view of Küçükyalı scenario}
    \label{fig:KucukyaliPlazaScenario1}
\end{figure}

To demonstrate the gains achieved by the RIS product in power-based and data rate-based communication quality metrics (e.g., RSRP, SINR, and UL/DL throughput), the RIS product has been integrated into the mmWave 5G infrastructure at Turkcell Küçükyalı Plaza to serve an end-user application. A virtual-reality (VR)-based ping pong application was established as part of this integration. The Google Earth view of the 5G mmWave wireless network scenario set up between the two buildings of Turkcell's main campus is provided in Fig. \ref{fig:KucukyaliPlazaScenario1}.

%%%%%%%%%%%%%%%%%%%%%%%%%%%%%%%
%%% Küçükyalı Scenario Footage  %%
%%%%%%%%%%%%%%%%%%%%%%%%%%%%%%%
\begin{figure}[h]
    \centering
    \subfigure[from RIS to AAU]
    {\includegraphics[width=0.32\columnwidth]{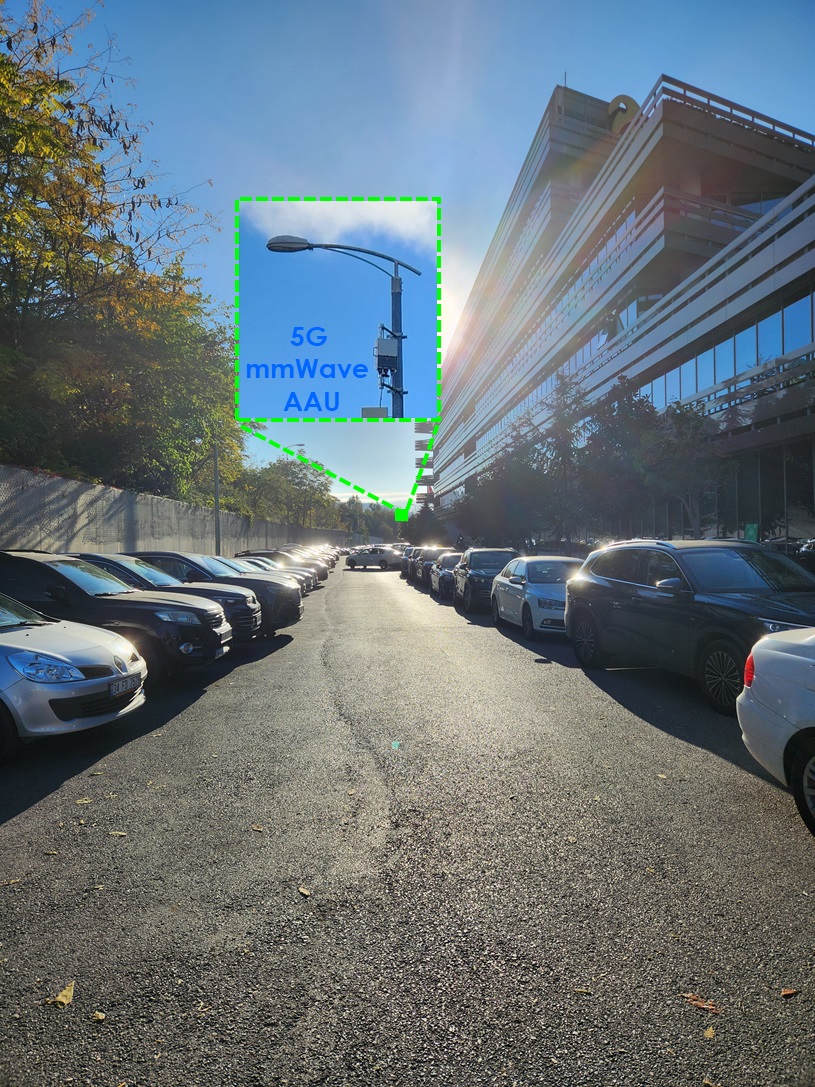}}
    \subfigure[from CPE to RIS]
    {\includegraphics[width=0.32\columnwidth]{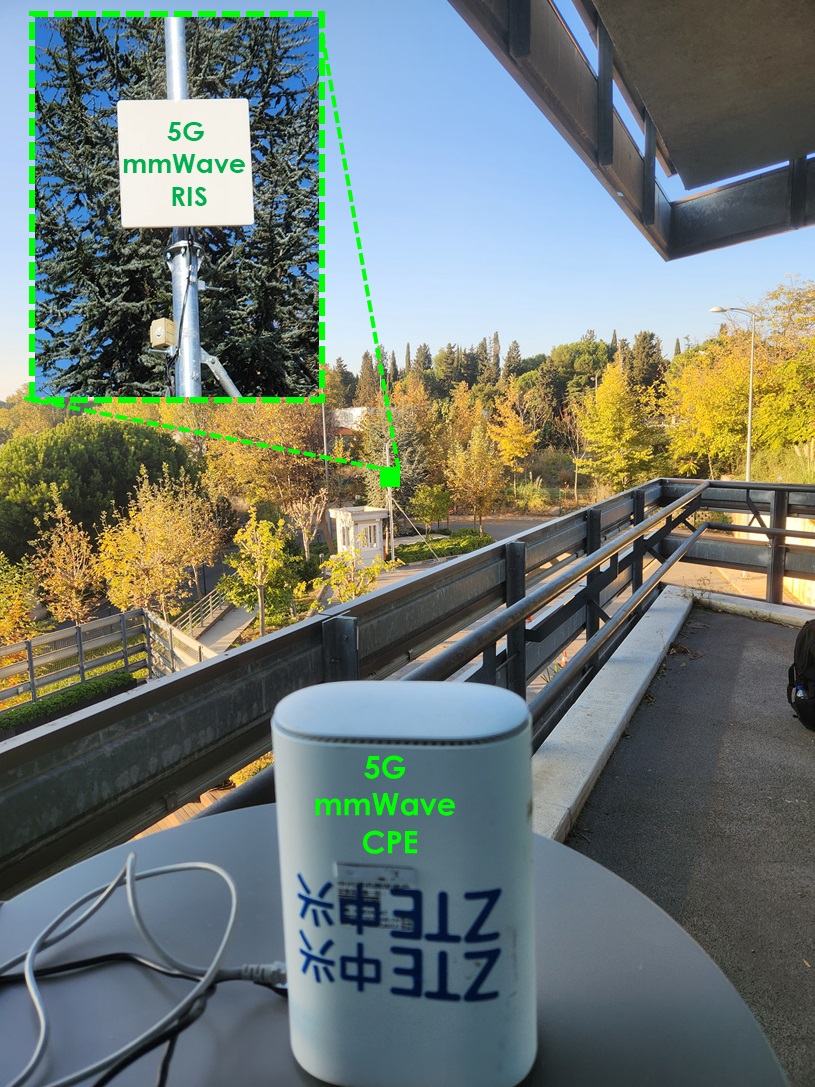}}
    \subfigure[from CPE to VR]
    {\includegraphics[width=0.32\columnwidth]{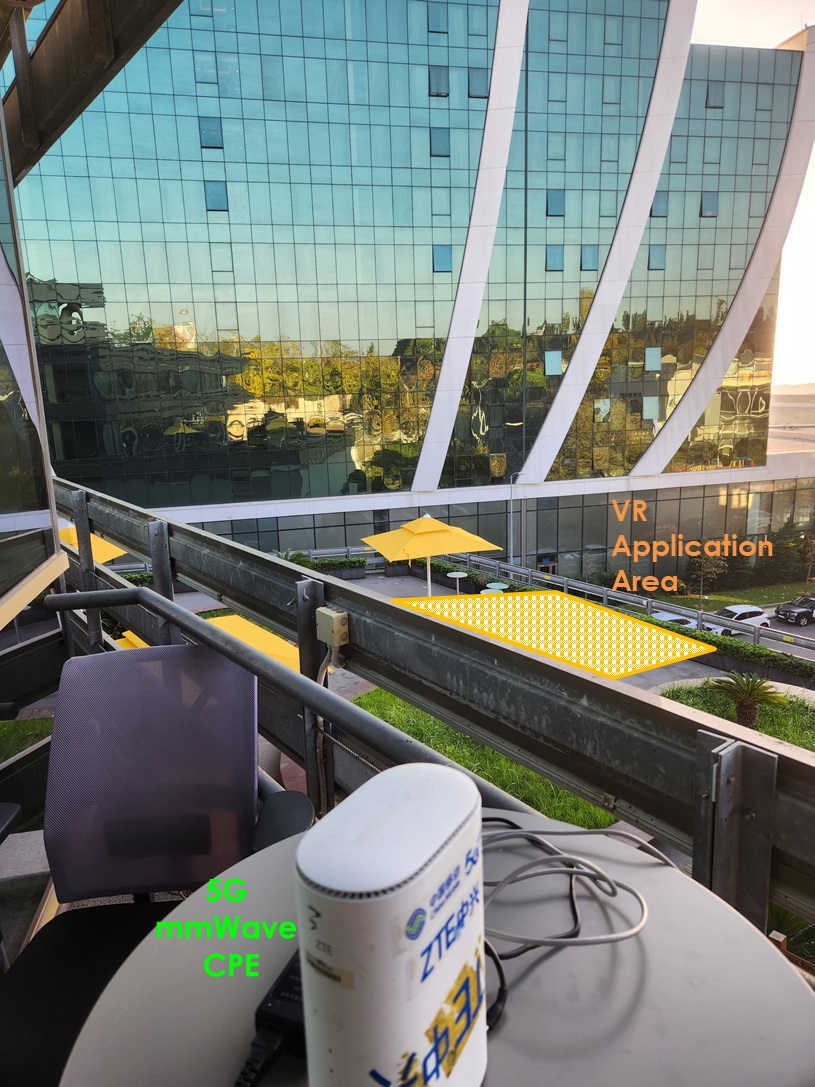}}    
    \caption{Footages of Küçükyalı Scenario}
    \label{fig:KucukyaliScenarioFootages}
\end{figure}

\begin{figure*}[h]
    \centering
    \subfigure[Original footage]
    {\includegraphics[width=0.98\columnwidth]{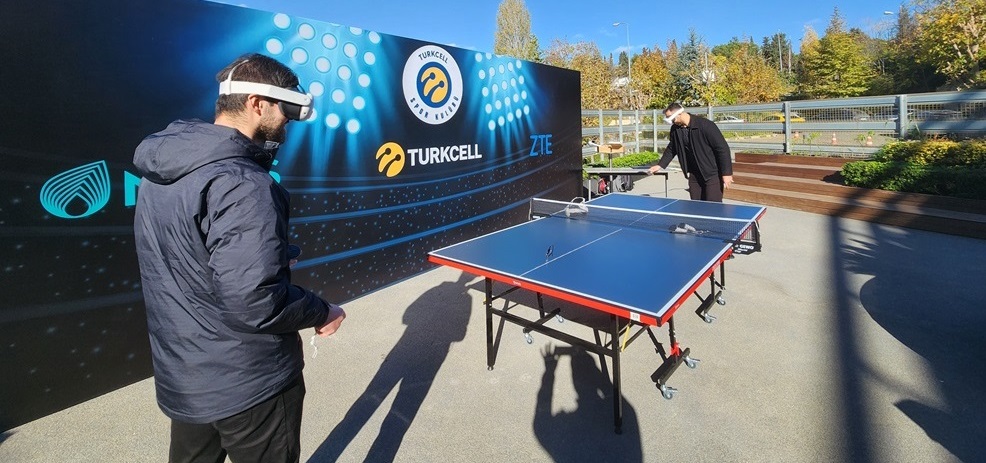}}
    \subfigure[VR game screen]
    {\includegraphics[width=0.98\columnwidth]{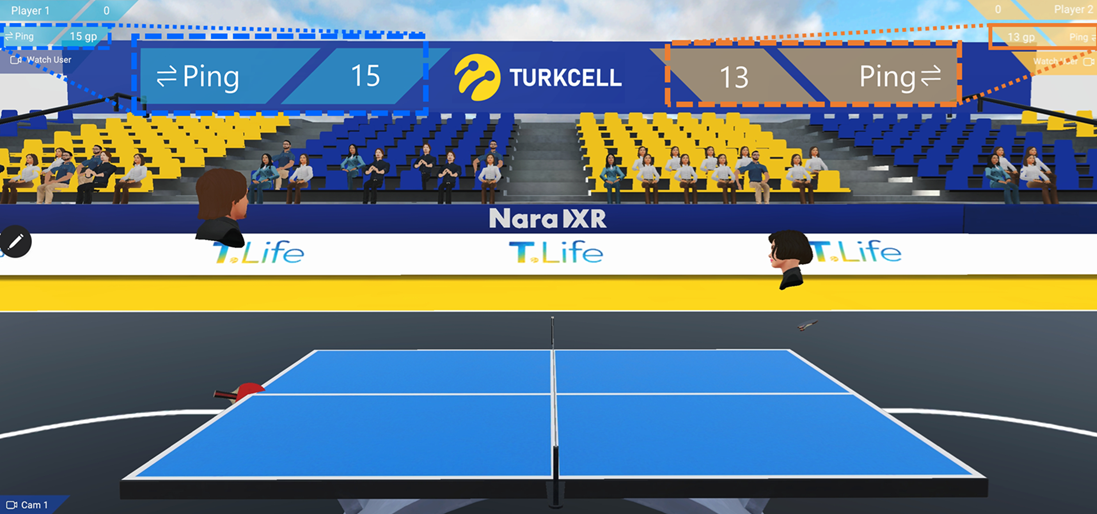}}    
    \caption{RIS-assisted VR-based game}
    \label{fig:KucukyaliVRGame}
\end{figure*}

%%%%%%%%%%%%%%%%%%%%%%%%%
The geographical information of the network equipment used in the VR application demo is given in Table \ref{table2:GeoInfoKucukyali}.  
%%%%%%%%%%%%%%%%%%%%%%%%%
\begin{table}[h]
  \centering
  \caption{\textsc{Geographical Information}}
  \label{table2:GeoInfoKucukyali}
  \begin{tabular}{c c c c c}
    \hline
    {\bf{Equip.}} & {\bf{Latitude [$^\circ$]}} & {\bf{Longitude [$^\circ$]}} & {\bf{Altitude [m]}} & {\bf{$d_{RIS}$ [m]}}\\
    \hline
    \hline
    {\bf{AAU: A}} & $40.955490$ & $29.118464$ & $74.2$ & $175.49$\\\hline
    {\bf{RIS: R}} & $40.956396$ & $29.116752$ & $70.8$ & $-$\\\hline
    {\bf{CPE: C}} & $40.956091$ & $29.116863$ & $68.8$ & $35.15$\\\hline
  \end{tabular}
\end{table}
In the mmWave RIS-assisted scenario with given locations, the RIS deployed at a corner point with a direct LoS to the 5G AAU aims to provide a high-bandwidth and low-latency mmWave connection to the CPE device located in the terrace area of Block C, which cannot be solely covered by the AAU. Since the VR devices used for the application are not inherently mmWave-compatible and are designed to establish data connections via a WiFi interface, the mmWave/WiFi6 conversion capability of the CPE has been utilized. Thus, the high-quality mmWave connection provided by the RIS is converted into a WiFi signal by the CPE and directed to the VR-based gaming area labeled as 'VR application area' in Fig. \ref{fig:KucukyaliPlazaScenario1}. The footages of the Küçükyalı demo scenario have been given in Fig. \ref{fig:KucukyaliScenarioFootages}. 
%%%%%%%%%%%%%%%%%%%%%%%%%%%%%%%
%%% Küçükyalı Demo Footages  %%
%%%%%%%%%%%%%%%%%%%%%%%%%%%%%%%
%%%%%%%%%%%%%%%%%%%%%%%%%
The investigated mmWave RIS-assisted VR game scenario is composed of a WiFi-based VR game and a RIS-assisted mmWave link that is more like a fixed wireless access (FWA) case different from Kartal Plaza trials. For this FWA scenario, the experienced RSRP, DL Throughput, and the end-to-end (E2E) latency in the absence and presence of RIS have been given in Table \ref{table3:KucukyaliKPIs}. 
%%%%%%%%%%%%%%%%%%%%%%%%%
\begin{table}[h]
  \centering
  \caption{\textsc{Experienced KPIs}}
  \label{table3:KucukyaliKPIs}
  \begin{tabular}{c c c c c}
    \hline
    {\bf{RIS State}} & {\bf{RSRP [dBm]}} & {\bf{DL Thru. [Mbps]}} & {\bf{E2E Latency [ms]}}\\
    \hline
    \hline
    {\bf{Off}} & $-98.28$ & $168.22$ & $>40$\\\hline
    {\bf{On}} & $-74.73$ & $1550.84$ & $[11,22]$\\\hline
  \end{tabular}
\end{table}
Note that the E2E latency denotes the aggregated latency including the signal round trip times (RTT) in mmWave and WiFi6 waveforms and the VR-based server processing where the developed ping pong game application has a typical constraint of $35$ ms E2E latency for stable operation. Since WiFi6 is mentioned to be suitable for VR-based applications with its RTT within the interval $[8,20]$ ms \cite{ZTEWiFi6}, and the VR-based server processing duration is typically $5$ ms, to satisfy the E2E latency constraint of the VR-based game, the mmWave link should guarantee typical marginal RTT lower than $10$ ms. Hence, switching RIS on or off has caused the latency requirement to be satisfied or violated, consequently directly affecting the VR-based user experience. When the RIS is on, the resulting gaming service is shown to function successfully, while E2E latency varies within the interval of $[11,22]$ ms. Fig. \ref{fig:KucukyaliVRGame} provides a real footage from the VR gaming area, and a snapshot from VR game interface showing the E2E latencies of both players. As the RIS is switched off, the E2E latencies drastically increase which results in corruptions in the gaming experience.

%%%%%%%%%%%%%%%%%%%%%%%%%
\section{Discussions}
Although results demonstrate that RIS technology possesses large potential to enhance future communicaiton networks, there are significant issues that need further investigation before they can be deployed in real-life scenarios.

\subsection{OSS Integration of RIS products}
Among these is the integration of RIS products with operational support systems (OSS). ZTE Dynamic RIS 2.0 and other current prototypes function as separate network components without OSS integration, hence limiting their manageability and scalability in live networks. Providing RIS solutions and set-ups that are compatible with OSS is crucial to allowing fault management, dynamic reconfiguration, and real-time monitoring.

\subsection{Multi-user extension}
The RIS in this study is only able to serve one user at a time, which limits the practicality in crowded urban settings, hence, multi-user scenarios poses another challenge. Further studies should investigate advanced techniques for RIS-based multi-user beam management and resource allocation, which can allow multiple users to be served simultaneously without performance degradation. 

\subsection{Interference management}
Another major issue involves the way RIS operations influence nearby networks and other users. In conjunction with the RIS supported network, the RIS introduces additional reflective paths that may disrupt other networks or connections. Extensive evaluations of these interference dynamics are essential for RIS-enabled networks to coexist in shared environments. Such challenges illustrate how further research and development is required in order to ensure effective implementation of RIS technology in realistic networks.

\section{Conclusions and Future Works}
In this study, to demonstrate the potential of the Dynamic 2.0 RIS prototype in enhancing network performance,  diverse real-world scenarios have been conducted in Turkcell's mmWave infrastructure. Through the tests in urban areas, significant improvements in mmWave coverage have been observed. In addition, the RIS-aided multi-user efficiency has been evaluated with two user equipment. Furthermore, a VR application has been integrated into the RIS-assisted communications link, fulfilling the low-latency requirements necessary for the smooth and immersive VR experience. Building on the findings, future studies will focus on evaluating multi-user interference and multi-BS configurations to evaluate the RIS performance in more complex scenarios, aiming to optimize RIS-aided systems for dense urban environments.

\section*{Acknowledgment}
This study has been supported by the 1515 Frontier Research and Development Laboratories Support Program of T{\"U}B{\.I}TAK under Project 5229901 - 6GEN. Lab: 6G and Artificial Intelligence Laboratory. 
The authors express their gratitude to ZTE Corporation for sharing the KPI results from outdoor tests, which paved the way to evaluate the RIS-based enhancements, and to NaraXR for their VR-based software development activities to provide an end-user application for the RIS-assisted mmWave network deployment. Special thanks are extended to our colleagues from ZTE, Tang Hu Jie, Zhang Ting, and Ataberk Kacargil for their invaluable contributions.

\bibliographystyle{IEEEtran}

\end{document}